\documentclass[a4paper,11pt]{article}
\usepackage{multirow}
\usepackage{mathptmx,amssymb}
\usepackage{algorithm}
\usepackage{algorithmic}
\usepackage{natbib}
\usepackage{graphicx}
\usepackage{latexsym,amsfonts,amsmath,amssymb}

\usepackage{fullpage}

\begin{document}

\title{A Bayesian approach to estimate changes in condom use from limited HIV prevalence data}

\author{J. Dureau$^1$, K. Kalogeropoulos$^1$, P. Vickerman$^2$,   M. Pickles$^3$, MC. Boily$^3$\\
\smallskip\\
{\it 1: London School of Economics,} \\
{\it Statistics Department, London, UK.}\\
\smallskip\\
{\it 2:  London School of Hygiene and Tropical Medicine, }\\
{\it Department of Global Health and Development, London, UK.}\\
\smallskip\\
{\it 3: Department of Infectious Diseases Epidemiology,}\\
{\it School of Public Health, Imperial College, London, UK.}}

\maketitle

\begin{abstract}
Evaluation of HIV large scale intervention programmes is becoming increasingly important, but impact estimates frequently hinge on knowledge of changes in behaviour such as the frequency of condom use (CU) over time, or other self-reported behaviour changes, for which we generally have limited or potentially biased data. We employ a Bayesian inference methodology that incorporates a dynamic HIV transmission dynamics model to estimate CU time trends from HIV prevalence data. Estimation is implemented via particle Markov Chain Monte Carlo methods, applied for the first time in this context. The preliminary choice of the formulation for the time varying parameter reflecting the proportion of CU is critical  in the context studied, due to the very limited amount of CU and HIV data available. We consider various novel formulations to explore the trajectory of CU in time, based on diffusion-driven trajectories and smooth sigmoid curves. Extensive series of numerical simulations indicate that informative results can be obtained regarding the amplitude of the increase in CU during an intervention, with good levels of sensitivity and specificity performance in effectively detecting changes. The application of this method to a real life problem illustrates how it can help evaluate HIV intervention from few observational studies and suggests that these methods can potentially be applied in many different contexts.
\end{abstract}

\section{Introduction}
Significant resources are being committed to implement large-scale interventions against infectious diseases, such as HIV/AIDS that killed an estimated two million individuals in 2008 \citep{Unaids2009}. Although such interventions are implemented on a large scale because they are expected to work, increasing attention is given to the evaluation of these large-scale intervention programmes to understand what still needs to be done to control the epidemic and eventually achieve elimination, ensuring that resources are not wasted on strategies that do not work.

Even if antiretroviral therapy has become an important component of large scale prevention interventions, condom use and circumcision remain important strategies for reducing HIV transmission. While there are difficulties in estimating condom use trends accurately, due to biases inherent in self-reported behaviour \citep{Turner1997,Zenilman1995,Hanck2008}, its average level closely determines the spread of HIV \citep{Boily2007}. Thus, it is important to assess if trends in epidemiological data such as HIV prevalence can be used to infer the impact of interventions on risk behaviours that are susceptible to self-reported bias. This is motivated by the fact that directly observed quantities as HIV prevalence do not provide straightforward indications on the impact of an intervention. Indeed, an epidemic has an intrinsic dynamic, which can cause the prevalence to grow although an efficient intervention is being led if the intervention is introduced early in an epidemic. Alternatively, in a mature epidemic the prevalence can decrease even though on-going interventions are inefficient \citep{Boily2002}. However, the trajectory of CU over time, and especially since the beginning of a prevention programme, can shed light on the impact of the intervention and on the future trajectory of the epidemic. In this light, we apply a Bayesian methodology to trends in HIV prevalence data, focusing on the specific example of Avahan, India AIDS initiative, a large-scale HIV/AIDS intervention targeted to high risk groups.

The Avahan intervention was motivated by high levels of HIV prevalence amongst high-risk groups observed in southern India (typically $>20\%$) (Ramesh, et al., 2008), which lead to concerns about infections bridging to their long-term partners and the general population. The programme was launched by the Bill \& Melinda Gates Foundation in 2003 (BMGF, 2008), and has targeted high-risk groups for HIV infection, in particular female sex workers (FSWs), by promoting and distributing free condoms. Different studies have been conducted to examine the impact of Avahan \citep{Boily2007,Deering2008,Boily2008,Lowndes2010,Pickles2010,Boily2013,Pickles2013}, and to learn from it in order to inform future large-scale interventions. A key part of such evaluations is examining how risk behaviours, chiefly condom use (CU), defined as the proportion of sex acts protected by condoms at a given time, have changed over the course of the intervention. However, this can be difficult to measure in practice. Baseline CU may be difficult to record when an intervention needs to be implemented rapidly, as happened with Avahan, or may be recorded only on few occasions. While those targeted by the intervention may be asked about their CU history \citep{Lowndes2010}, their answers may be subject to social desirability and recall biases. In principle, the total number of condoms sold or distributed can be enumerated \citep{Bradley2010}, but accurate records may not be available, condoms may be used for family planning by lower-risk individuals, and the distribution of condoms is not a guarantee of their correct usage \citep{Bradley2010,Kumar2011}. Thus, in addition to direct approaches through quantitative behavioural surveys or records of condom availability, model-based methods can be used to infer unobserved quantities of interest, such as CU, and complete the partial information available from observed quantities such as HIV prevalence, using knowledge of the dynamics of large-scale epidemics.

A first study in the context of Avahan was presented in \cite{Pickles2010}, \cite{Boily2013} and \cite{Pickles2013}. In this work, a deterministic dynamic model for HIV/sexually transmitted infection was formulated based on a compartmental representation incorporating heterogeneous sexual behaviour. The main aim of this approach was to shed light upon the partially observed trajectory of CU based on different data sources, with their own potential biases. This initial inference proceeded by setting three different hypothesised scenarios for the CU evolution over time, before and after the introduction of Avahan in 2003, that were estimated using either self reported CU data from a number of serial cross-sectional surveys on FSWs \citep{Lowndes2010} or using condom distribution and condom sales data. The three hypotheses alternatively assumed 
\begin{enumerate}
\item CU initially increased according to the reported FSWs survey data until the Avahan initiative started in 2003 and was assumed to remain constant afterwards, despite evidence of an increase from FSW survey data. This was to acknowledge that FSWs may over-report condom use after being exposed to the intervention.
\item CU increased as in scenario (a) up to the Avahan initiative, but CU was assumed to increase from 2003 onwards in accordance with self reported FSWs survey data.
\item Same as scenario (b) but the increase after 2003 occurred according to condom distribution/sales data.
\end{enumerate}
The HIV/sexually transmitted infection model was then used with each of the above scenarios, which were therefore assessed based on the fit to the available prevalence observations.

The work we present in this article operates in a similar context as in \cite{Pickles2010}, but rather than making explicit assumptions about the evolution of CU based on additional sources, we aim in developing a Bayesian inference framework exploring the entire space of the CU trajectories without any reference to the 2003 Avahan intervention, FSWs surveys nor condom distribution data. This enables us to assess the impact of the Avahan intervention on CU under less tangent assumptions, thereby providing additional and stronger evidence of the impact of Avahan on CU, beyond the three scenarios above. From a methodology perspective, the model formulation of our approach can be put in a state space setting where an underlying latent process (CU trajectory) is observed through the prevalence data, and the link between these quantities is given by the deterministic model for the HIV infections. Inference in this context is a challenging task given the limited amount of HIV prevalence data aside from initial conditions (three or four observations in total) that are concentrated over a period of 6 years, and are utilised to estimate a 25-years long trajectory. Various models for the CU trajectories were considered, including smooth and non-differentiable (yet continuous) choices. In the remainder of this paper, the term `trajectory prior' is used to refer to these models in order to avoid confusion with the deterministic HIV model. Note however that the trajectory priors include parameters for which there exists some information in the data. We also present a general and efficient computational scheme using Markov Chain Monte Carlo (MCMC) techniques based on the particle MCMC algorithm (\cite{Andrieu2010}; see \cite{Dureau2012} for an application in a different context). Focus is given on estimating the amplitude of the change in CU since 2003 (the start of Avahan) in order to assess the impact of the Avahan intervention on CU. In addition to the scientific interest of this application, there are various methodological challenges given the limited amount of data. For this reason, the properties of the estimators arising from the MCMC are studied via simulations, and the performance is assessed from a decision-making perspective through their sensitivity and specificity in detecting changes in CU.

The next section presents the models introduced in this paper, the data that are typically available for such studies, and the way in which prior information is incorporated. The computational techniques, mainly the particle MCMC algorithm, are also presented. The developed methodology to compare the performance of the proposed trajectory priors is presented in Section \ref{EvalMeth}, and the results from this study are introduced in Section \ref{Results} along with an application to real data from the Indian AIDS initiative Avahan. Finally Section \ref{Discussion} concludes with some relevant discussion.

\section{\label{ModelsMethods}Models and Methods }

Our modelling specification consists of two parts. The first part is based on the formulation of \cite{Pickles2010}, for the HIV transmission population dynamics. Our proposed methodology targets mainly the second part where the adopted HIV transmission model is extended by considering a parametric time varying version of the CU parameter. In what follows we first introduce and describe the HIV transmission model (section \ref{subHIVmodel}), before getting into our proposed extension of it (section \ref{subseqTrajPriors}), its prior parameterisation (\ref{subseqPriors}) and its MCMC implementation (\ref{subseqCompScheme}).

\subsection{\label{subHIVmodel}HIV transmission model for female sex workers}
\begin{figure}

    \centering
    \makebox{\includegraphics[width=10cm]{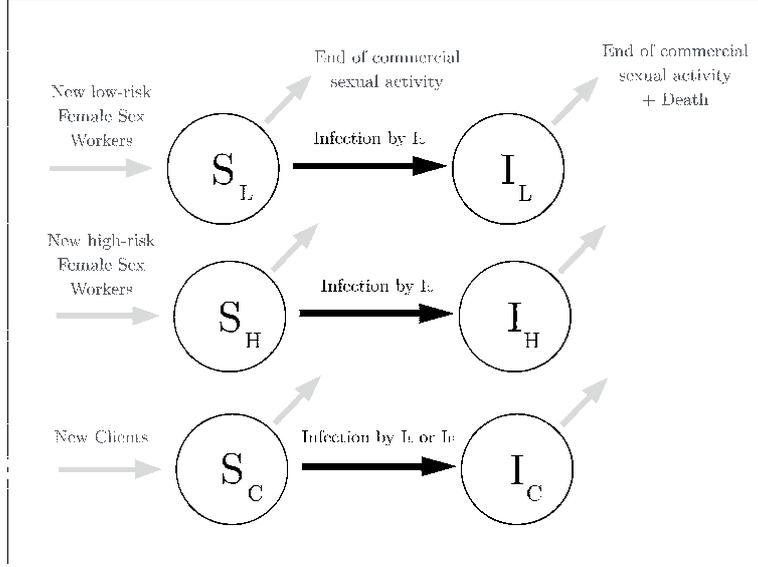}}
    \caption{Flow-diagram of the model.}
    \label{fig:modelflowchart}
  \end{figure}

In this section we present the part of our modelling framework that aims to capture the dynamics of HIV transmission in an environment with features as in the Mysore district in Southern India. The model is based on that of \cite{Pickles2010} and can be described as follows: a stable but open population is considered, consisting of FSWs and their clients. In line with \cite{Vickerman2010}, we assume that the remaining members of the population (eg. companions of clients or sex workers), that are not involved directly in sex work, have little influence on the dynamics of the epidemic and can therefore be ignored. The model structure allows for two different groups of FSWs, i.e. high-risk ($H$) and low-risk ($L$), that may have different numbers of clients ($C$). Each individual in these three groups can either be susceptible to HIV infection, infectious, or retired either due to death or ceasing commercial sexual activity. Moreover, individuals that either decease or stop being involved in commercial sex are assumed to be replaced by susceptible ones, such that the size of the risk population remains constant. A graphical representation of the model is given in Figure \ref{fig:modelflowchart} via a flow-diagram for each of these three population groups.
A key component of the epidemic model is the force of infection $\beta(t)$ which is assumed to be the product of the number of clients multiplied by the probability of HIV transmission in at least one out of $n$ sexual acts per client encounter. A sexual act can lead to HIV transmission when either no condom is being used, occurring with probability $1-CU(t)$, or when a condom is not being used effectively, occurring with probability $CU(t)(1-e)$. Hence, the combined probability of a sexual act that can lead to HIV transmission, also referred to as risky sexual act, simplifies to $1-eCU(t)$. The probability of HIV transmission from the sex worker to the client (or from the client to the sex worker) in a single risky sexual act is given by $p_S$ (or $p_C$), thus providing the expression of the force of infection; e.g. for high risk FSWs we get $C_H[1-\{1-p_{S}\times(1-e CU(t))\}^{n}]$, where $C_H$ is the number of clients for high risk FSWs:
%
\begin{equation}
\begin{array}{rl}
\beta_{H}(t) & =C_{H}[1-\{1-p_{S}\times(1-e CU(t))\}^{n}]\\
\beta_{L}(t) & =C^{L}[1-\{1-p_S\times(1-e CU(t))\}^{n}]\\
\beta_{C}(t) & =\frac{C_H+C_L}{2}\frac{N_{F}}{N_{C}}[1-\{1-p_{C}\times(1-e CU(t))\}^{n}]
\end{array}
\label{eq1}
\end{equation}
where the following notations are used:
\begin{itemize}
\item $C_H$ and $C_L$: number of clients of FSWs per month, which differs for high risk and low risk FSW
\item $n$: number of acts per client encounter
\item $p_S$ or $p_C$: probability of HIV transmission from client to sex worker or sex worker to client respectively during an unprotected sex act
\item $e$: efficacy of condoms in protecting against transmission of HIV per sex act
\item $CU(t)$: proportion of FSWs' sex acts  that are protected by condoms, that we allow to vary in time (it is the main parameter that we want to estimate)
\end{itemize}
Additionally, the transmission dynamics of HIV will depend on the following lengths of time:
\begin{itemize}
\item  $\mu_S^{-1}$ or $\mu_C^{-1}$: average length of sexual activity as a sex worker / client
\item	 $\alpha^{-1}$:  average life expectancy with HIV
\end{itemize}
Information on the above quantities is available from various serial cross-sectional bio-behavioural surveys (IBBAs) in Mysore district in southern India (Ramesh, et al., 2008). However, there is still uncertainty around these biological and behavioural parameters, which is reflected on the estimates of CU by using a Bayesian approach \citep{DeAngelis1998}.

The model can also be defined via the following set of differential equations:
\begin{equation}
\left\{
\begin{array}{rl}
\frac{dS_{H}}{dt}(t) & = -\beta_{H}(t)S_{H}(t)\frac{I_{C}(t)}{N_{C}} + (\mu_{S}+\alpha) I_{H}(t)\\
\frac{dI_{H}}{dt}(t) & =\;\;\;\beta_{H}(t)S_{H}(t)\frac{I_{C}(t)}{N_{C}}-(\mu_{S}+\alpha)I_{H}(t)\\
\frac{dS_{L}}{dt}(t) & =-\beta_{L}(t)S_{L}(t)\frac{I_{C}(t)}{N_{C}} + (\mu_{S}+\alpha) I_{L}(t)\\
\frac{dI_{L}}{dt}(t) & =\;\;\;\beta_{L}(t)S_{L}(t)\frac{I_{C}(t)}{N_{C}}-(\mu_{S}+\alpha)I_{L}(t)\\
\frac{dS_{C}}{dt}(t) & =
  -\beta_C(t)S_{C}(t)\left\{\frac{C_HN_{H}}{C_HN_{H}+C_LN_{L}}(\frac{I_{H}(t)}{N_{H}})+\frac{C_LN_{L}}{C_HN_{H}+C_LN_{L}}(\frac{I_{L}(t)}{N_{C}})\right\}\\&\;\;\;\;\;\; + (\mu_{C}+\alpha) I_{C}(t)\\
  \frac{dI_{C}}{dt}(t)  & =  \;\;\;\beta_C(t)S_{C}(t)\left\{\frac{C_HN_{H}}{C_HN_{H}+C_LN_{L}}(\frac{I_{H}(t)}{N_{H}})+\frac{C_LN_{L}}{C_HN_{H}+C_LN_{L}}(\frac{I_{L}(t)}{N_{L}})\right\}\\ & \;\;\;\;\;\;-(\mu_{C}+\alpha)I_{C}(t)
\end{array}\right.
\label{eq2}
\end{equation}

In the equations above, the state of the epidemic over time is described by the number of susceptibles among clients ($S_C$), among the low-risk female sex workers ($S_{L}$) and  among the high-risk female sex workers ($S_{H}$), as well as the corresponding numbers of infected individuals ($I_C$, $I_{H}$ and $I_{L}$). Three types of constant parameters are involved: the initial prevalence among the different groups of interest in 1985 ($\theta_{i.c.}= \{S_{H}(1985), I_{H}(1985), S_{L}(1985), I_{L}(1985), S_{C}(1985),  I_{C}(1985)$), time-invariant parameters describing the biological and behavioural determinants of HIV transmission ($\theta_{tr.}=\{\mu_S,\mu_C,\alpha,C_H, C_L,p_S ,p_C  ,n,e$\}) and the parameters that play a role in the trajectory priors for CU between 1985 and 2010 ($\theta_{CU}$). All three components $\theta_{i.c.},\theta_{tr.}$, and $\theta_{CU}$ are integrated into a global vector of constant parameters, denoted by $\theta$. Under this notation, the trajectory $X_{1:n}$ of the space vector $X(t)=\{ S_{H}(t), I_{H}(t), S_{L}(t),$ $ I_{L}(t), S_C(t),  I_C(t)\}$ is defined as a deterministic function of $\theta$ and CU ($X_{1:n}=f(\theta,CU)$) through an HIV transmission model, and is compared with the available observations,  denoted by  $y_{1:n}$. Note that the function $f(.)$ is not available in closed form but can be obtained given the trajectory of $CU$ by solving the above ordinary differential equations (ODE). More specifically, we introduce a time discretisation with equidistant points of time step $\delta$ resulting in a discretised skeleton of CU denoted $CU^{discr} =\{CU(t_0) ,CU(t_0+\delta) ,CU(t_0+2\delta) ,..,CU(t_n)  \}$. The partition of the CU trajectory can be made arbitrarily fine by the user-specified parameter $\delta$ to limit the approximation error induced by the time discretisation.

Assigning a model for the observation error provides the likelihood of the observation $y_{1:n}$ conditional on the CU trajectory $p(y_{1:n}| \theta, CU)$. In this paper we use a binomial distribution, considering that prevalence estimates are derived from a random sample of 425 FSWs or clients in the Mysore district. We denote $h_i(X)$ the value of the prevalence estimated through the HIV transmission model at time $i$: $h_i\{X(t_i)\}=\{I_{H}(t_i)+ I_{L}(t_i)\}/2$ when prevalence among FSWs is observed,  $h_i\{X(t_i)\}=I_C(t_i)$ if prevalence among clients is observed. With these notations, the likelihood function is defined from the following:
\begin{equation}
425\times y_i \sim  Bin\{425,h_i[X(t_i)]\}\nonumber
\end{equation}
Overall, the model appearing in Figure \ref{fig:modelflowchart} and Equations \ref{eq1} and \ref{eq2} is a simplified version of the one in (Pickles, et al., 2010). This was done mainly for parsimony reasons; models of increased complexity can be used provided that there is adequate information on their parameters. More details on informative priors are provided in Section \ref{subseqPriors}.

\subsection{\label{subseqTrajPriors}Trajectory priors for condom use}

From a modelling perspective, the main contribution of this paper lies in the introduction of various formulations for the evolution of the CU, aiming to explore its space of trajectories rather than evaluate a limited set of scenarios. The proposed trajectory priors are motivated by two different objectives. First we assign a trajectory prior aiming to impose little information to its shape. We proceed with a Brownian motion for CU, transformed to take values in the real line, although integrals of Brownian motion can also be used. Loosely speaking these models can be linked with a smoothing splines approach \citep{Wahba1990}. The second objective aims for a slightly more informative formulation regarding the evolution of CU via sigmoid-shaped growth curves. Growth curves provide natural models for smoothly growing quantities in different contexts such as biology \citep{Zwietering1990}, marketing \citep{Lessne1988} and epidemiology \citep{Omran1971}, while they were also mentioned in \cite{Pickles2010}. Moreover, the posterior of the CU trajectory, based on the model with Brownian motion as a prior, revealed a sigmoid shaped growth. Hence, the second trajectory prior, denoted by dBR, is based on the generalised Bertalanffy-Richards model; see for example \citep{Garcia1983,Yuancai1997}. In order to enrich this context and address estimation issues that can be encountered with the dBR \citep{Lei2004}, we also consider an alternative empirical sigmoid curve (dSigm). In what follows, we denote with $x$ the latent process that drives the CU trajectory, which in turn provides the link with the prevalence observations through the model in Section \ref{subHIVmodel}.

\subsubsection{Brownian motion (BM)}

The first formulation assigns a Brownian motion to a transformed version of the CU trajectory. As the latter has to be constrained in the [0,1] region, we work with the logit transformation of $CU(t)$, denoted by:
\begin{eqnarray}
CU(t) &=& \frac{\exp(x(t))}{1+\exp(x(t))}\nonumber\\
dx(t) &=& \sigma dB_t\label{eq:xBM}
\end{eqnarray}
The use of diffusion processes to describe time varying quantities in contexts associated with uncertainty has been used in epidemic models; see for example \cite{Cazelles1997}, \cite{Cori2009} and \cite{Dureau2012}. It can also be seen as a prior according to which $x(t)$  is a random walk with continuous, yet non-differentiable trajectories. It is used here in an attempt to incorporate a limited amount of prior information on the shape of the trajectory. It can also be used as an exploration tool for potential modelling-remodelling steps towards more informative formulations. Variations of this formulation may include smoother diffusion models, by taking integrals of the Brownian motion, or alternative transformations such as the probit link. We note at this point that very little information is available on the volatility in \eqref{eq:xBM} which is determined mostly by its prior. More details are provided in Sections \ref{subseqPriors} and \ref{EvalMeth}.

\subsubsection{Deterministic Bertallanfy-Richards function (dBR )}

Qualitatively, CU trends reconstructions by alternative methods \citep{Lowndes2010,Bradley2010} suggest that CU was quite low in 1985, and has grown over the recent year. The above motivated the use of a growth curve parametric model instead of the Brownian motion diffusion. We use the generalised Bertalanffy-Richards (BR) family \citep{Richards1959,Garcia1983} that can be written as:
\begin{equation}
CU(t)=\eta(1-Be^{-kt})^{\frac{1}{1-m}}\nonumber
\end{equation}
or else, in differential equation framework:
\begin{eqnarray}
\label{eq:CUBR}CU(t) &=& [(1-m)x(t)+\eta^{1-m}]^{\frac{1}{1-m}}\\
\label{eq:xBR}dx(t)  &= &-kx(t) dt
\end{eqnarray}
This family contains various growth curves, including the logistic ($m=2$) and Gompertz ($m\rightarrow \infty$) functions. The growth curve can be parameterised by four quantities: the initial value of  $CU_{0}$, the time of inflection $t_{in}$ (at which $CU''(t)=0$), the value of CU after an infinite time ($\eta$, also termed as the asymptote), and the `shape' or `allometric' parameter $m$. Note that the time of inflection can be related to the parameter $k$ by the following equation:
\begin{equation}
k\times t_{in} = \log(\frac{B}{1-m})\nonumber
\end{equation}
Furthermore, this definition implies that the initial value $CU_0$  is lower than $m^{\frac{1}{1-m}}\eta$. In order to focus on sigmoid-shaped growth curves, we restrict our attention to cases where $m\geq1$  \citep{Yuancai1997}.

\subsubsection{Deterministic empirical sigmoid curve (dSigm)}

An empirical sigmoid model is also considered to address the potential difficulties that can arise with the parameterisation of the dBR. Since growth models are used to study intrinsically growing objects, trajectories that are inexplicably stable for a long period of time and that eventually start picking at a rapid pace are not typical under the BR formulations. Moreover, inference on the allometric parameter $m$ in dBR can be problematic \citep{Lei2004}. These may lead to underestimating the amplitude of a shift in CU under the potential extrinsic influence of the Avahan intervention. For this reason, we also consider an alternative sigmoid curve, defined in the following way:
\begin{eqnarray}
CU(t) &=& a+\frac{b}{1+cx(t)}\nonumber\\
dx(t) &= &-kx(t) dt\label{eq:xSigm}
\end{eqnarray}
Here the model is parameterised by its baseline ($CU_0$ ), its asymptote ( $\eta$), its time of inflection  ($t_{in}$), and the increase rate ($k$), from which a, b and c can be computed:
\begin{eqnarray}
a &= & \eta - b\nonumber\\
b &= & (1+\frac{1}{c})\times(\eta-CU_0)\nonumber\\
c &=  & e^{kt_{in}}\nonumber\\
\end{eqnarray}

\subsubsection{Stochastic growth curves}

It is also possible to combine the Brownian motion and the growth curve approaches using diffusions. Stochastic extensions of the dBR and dSigm model can be considered, in which the mean behaviour remains intact while some random perturbations are introduced through a stochastic differential equation. In order to ensure positivity, restrict $CU(t)$ below one and retain the link with deterministic dBR curve, a geometric Brownian motion can be used to replace equations \eqref{eq:xBR} and \eqref{eq:xSigm}
\begin{equation}
\label{eq:Sgrowth}
dx(t) =-kx(t) dt + \sigma x(t) dB_t
\end{equation}
The stochastic growth curve defined by \eqref{eq:CUBR} and \eqref{eq:Sgrowth} was also mentioned in \cite{Garcia1983}. A convenient feature for both stochastic extension of dBR and dSigm is the fact that since
\begin{equation}
x(t)=\frac{1}{1-m}(CU(t)^{1-m}-\eta^{1-m}),\nonumber
\end{equation}
and $x(t)$ is strictly negative, the resulting CU trajectory is maintained strictly below $\eta$. Given the limited data at our disposal, these models can hardly be fitted in the context of this paper. Nevertheless, they may be helpful in cases where more observations are available.

\subsection{\label{subseqPriors}Priors}
The parameters contained in $\theta_{i.c.}$ and $\theta_{tr.}$ cannot be identified from the prevalence observations only, so we assign informative priors on them. These are summarised in Table \ref{Priors} and are similar with the priors used in \cite{Pickles2010}. They were either based on previous literature regarding general quantities as transmission probability for unprotected acts, or life expectancy with HIV, whereas the ones concerning quantities that are more sociologically and geographically specific were estimated from cross-sectional individual-based surveys (IBBAs) in Mysore.

The parameter vector $\theta$ includes an additional component, $\theta_{CU}$, that contains the parameters for different models describing the CU trajectories, $\theta = \{\theta_{i.c.},\theta_{tr.},\theta_{CU}\}$. Although there is some information in the data for $\theta_{CU}$, the posterior will depend on the prior to a large extent. As mentioned earlier, there is very little information on the volatility parameter of the BM formulation. Throughout this paper we used a Uniform prior between $0$ and $2$ $months^{-1}$. As explained in more detail in Section \ref{sec:DCU}, the parameter of main interest in this study is the quantity $\Delta CU= CU(2009)-CU(2003)$. Simulations suggest that, if we combine the BM approach with a $Unif(0,1)$ prior for $CU_0$ (CU in 1985), this results in a symmetric prior on $\Delta CU$ that is centered around 0 with 2.5\% and 97.5\% points at $\pm 0.6$ respectively. We considered it as a reasonably vague prior for $\Delta CU$ and evaluated the performance of the resulting model via the simulation experiments of Section \ref{EvalMeth}. More diffuse priors can also be used by setting a larger value for the upper limit of the Uniform prior for $\sigma$. Regarding the parameters of the sigmoid curves, we used vague priors that are also shown in Table \ref{Priors}.

\begin{table}
	\caption{\label{Priors}Table of priors for the different components of $\{\theta_{i.c.},\theta_{tr.},\theta_{CU}\}$}
\centering
\fbox{%
\begin{tabular}{|ccc|}
 \hline
 \hline
\emph{HIV transmission model } & \multirow{2}{*}{\emph{Notation}} & \emph{Range of uniform priors for the }\tabularnewline
\emph{parameters definition} &  & \emph{district of Mysore (Pickles et al. 2010)}\tabularnewline
\hline
\hline
Probability of transmission &  \multirow{2}{*}{$p_{S}$} &  \multirow{2}{*}{0.0006-0.0055} \tabularnewline
from Client to Sex Worker per act  & &\tabularnewline

Probability of transmission & \multirow{2}{*}{$p_{C}$} & \multirow{2}{*}{0.0001-0.007}\tabularnewline
from Sex Worker to Client per act &&\tabularnewline

 \multirow{2}{*}{Condom efficacy per act} &  \multirow{2}{*}{$e$} &  \multirow{2}{*}{80\%-95\%}\tabularnewline
 &&\tabularnewline

 \multirow{2}{*}{Mean number of acts per clients} &  \multirow{2}{*}{$n$} &  \multirow{2}{*}{1-2}\tabularnewline
  &&\tabularnewline

 \multirow{2}{*}{Mean number of clients per high-risk FSW} &  \multirow{2}{*}{$C_H$} &  \multirow{2}{*}{46.6-54.0 clients/month}\tabularnewline
  &&\tabularnewline

 \multirow{2}{*}{Mean number of clients per low-risk FSW} &  \multirow{2}{*}{$C_L$} &  \multirow{2}{*}{20-23.7 clients/month}\tabularnewline
  &&\tabularnewline

 \multirow{2}{*}{Toral number of FSWs} &  \multirow{2}{*}{$N_{H}+N_{L}$} &  \multirow{2}{*}{1943}\tabularnewline
  &&\tabularnewline

 \multirow{2}{*}{Cliens/FSW population ratio} &  \multirow{2}{*}{$\frac{N_{C}}{N_{H}+N_{L}}$} &  \multirow{2}{*}{7-19}\tabularnewline
  &&\tabularnewline

 \multirow{2}{*}{Mean length of sexual activity as FSW} &  \multirow{2}{*}{$\mu_{S}^{-1}$} &  \multirow{2}{*}{45-54 months}\tabularnewline
  &&\tabularnewline

 \multirow{2}{*}{Mean length of sexual activity as client }&  \multirow{2}{*}{$\mu_{C}^{-1}$} &  \multirow{2}{*}{154-191 months}\tabularnewline
  &&\tabularnewline

 \multirow{2}{*}{Mean life expectancy after infection with HIV} &  \multirow{2}{*}{$\alpha^{-1}$} &  \multirow{2}{*}{87-138.5 months}\tabularnewline
  &&\tabularnewline

 \multirow{2}{*}{Initial proportion of infected FSWs in 1985} &  \multirow{2}{*}{$\frac{I_{H}+I_{L}}{N_{H}+N_{L}}$} &  \multirow{2}{*}{0\%-5\%}\tabularnewline
  &&\tabularnewline

 \multirow{2}{*}{Initial proportion of infected clients in 1985} &  \multirow{2}{*}{$I_{C}/N_{C}$} &  \multirow{2}{*}{0\%-5\%}\tabularnewline
  &&\tabularnewline
\hline
 \hline
  \multirow{2}{*}{\emph{Condom trajectory priors parameters definition}} &  \multirow{2}{*}{\emph{Notation}} &  \multirow{2}{*}{\emph{Prior}}\tabularnewline
  &&\tabularnewline
 \hline
\hline
 \multirow{2}{*}{Allometric parameters (dBR)} &  \multirow{2}{*}{$m$} &  \multirow{2}{*}{$\mathcal{N}(1,10^{6})\times\mathbb{I}_{]1,+\infty[}$}\tabularnewline
  &&\tabularnewline

 \multirow{2}{*}{Growth rate (dSigm)} &  \multirow{2}{*}{$k$} &  \multirow{2}{*}{$\mathcal{N}(0,10^{6})\times\mathbb{I}_{]0,+\infty[}$}\tabularnewline
  &&\tabularnewline

 \multirow{2}{*}{Asymptote (dBR, dSigm)} &  \multirow{2}{*}{$\eta$} &  \multirow{2}{*}{$Unif(0,1)$}\tabularnewline
  &&\tabularnewline

 \multirow{2}{*}{Initial Value (all trajectory priors)} &  \multirow{2}{*}{$CU(t_{0})$} &  \multirow{2}{*}{$Unif(0,1)$}\tabularnewline
  &&\tabularnewline

 \multirow{2}{*}{Time of inflection (dBR, dSigm) }&  \multirow{2}{*}{$t_{in}$} &  \multirow{2}{*}{$Unif(1985,2009)$}\tabularnewline
  &&\tabularnewline

Allometric parameters, initial  & \multirow{2}{*}{$[CU(t_{0}),\eta,m]$} &\multirow{2}{*}{ 0 if $CU_{t_{0}}\geq m^{\frac{1}{1-m}}\eta$}\tabularnewline
conditions and asymptote  (dBR) &&\tabularnewline

 \multirow{2}{*}{Volatility (BM)} &  \multirow{2}{*}{$\sigma$} &  \multirow{2}{*}{$Unif(0,2)\;months^{-1}$}\tabularnewline
  &&\tabularnewline
\hline
\end{tabular}}
\end{table}

\subsection{\label{subseqCompScheme}Computational schemes for implementation }

The joint posterior distribution can be obtained up to proportionality through the HIV transmission model of Section \ref{subHIVmodel}, which links the prevalence observations with the CU trajectories, the trajectory priors of Section \ref{subseqTrajPriors} and the remaining priors of Section \ref{subseqPriors}. For the dBR and dSigm trajectory priors, it can be put in a non-linear regression framework, with the non-linear function being the solution of the ODE, and can therefore be implemented with standard software such as WinBUGS through WBDiff \citep{Lunn2004}. However this is not possible for the BM case where more involved techniques are required. Since the posterior probability density function is intractable, a data augmentation scheme can be utilised. This inference problem poses some challenges due to the high dimension of the discretised representation of $CU_{1:n}$ and its strong correlation with the vector of constant parameters,  $\theta$. This correlation imposes problems to Gibbs schemes on $\theta$ and $CU_{1:n}$, leading to extremely poor mixing and convergence properties. The Particle MCMC algorithm (PMCMC, see \cite{Andrieu2010}) algorithm offers a solution by updating the two components jointly, thus reducing the problem to a small-dimensional MCMC on $\theta$. Implementation is based on the estimates of the likelihood $\hat{p}(y|\theta)$ that are provided by a particle filter.

The particle filter and the PMCMC are described in Algorithms \ref{alg:SMC} and \ref{alg:PMCMC}, in terms of the quantities introduced in the previous sections. More details about this algorithm and its practical implementation can be found in \cite{Dureau2012} (through an application in a similar context) and \cite{Andrieu2010}.

\begin{algorithm}[h]
\caption{Particle Filter algorithm}
\label{alg:SMC}
\begin{algorithmic}
\STATE With $N$ being the number of particles and $n$ the number of observations.
\STATE Initialise $L^0(\theta)=1$, $W_{0}^{j}=\frac{1}{N}$, sample $(\widetilde{CU}(t_0)^{j})_{j=1,\dots,N}$ from $p(CU(t_0)|\theta)$
\FOR {$i=0$ to $n-1$}
	\FOR {$j=1$ to $N$}
		\STATE Sample $(\widetilde{CU}(t_i:t_{i+1})^{j})$ from $p(CU(t_i:t_{i+1})|\theta,CU(t_i)^j)$
		\STATE Calculate the resulting prevalence $h\{\widetilde{X}(t_{i+1})\}^{j, model}$  by solving the ODE (for example with the Euler step)
		\STATE Set $\alpha^{j}= p(y_{i+1}|h\{\widetilde{X}(t_{i+1})\}^{j,model})$
	\ENDFOR
	\STATE Set $W_{i+1}^{j}=\frac{\alpha^{j}}{\sum_{k=1}^{N}\alpha^{k}}$, and $L^{i+1}(\theta)=L^{i}(\theta)\times \frac{1}{N} \sum \alpha^{j}$
	\STATE Resample $(\widetilde{CU}(t_{1}:t_{i+1})^{j},\widetilde{X}(t_{i}:t_{i+1})^{j,model})_{j=1,\dots,N}$ according to $(W_{i+1}^{j})$,
\ENDFOR		
\end{algorithmic}
\end{algorithm}

\begin{algorithm}[h]
\caption{Particle MCMC algorithm (particle Marginal Metropolis Hastings version)}
\label{alg:PMCMC}
\begin{algorithmic}
\STATE With M being the number of iterations
\STATE Set current $\theta$ value, $\tilde{\theta}$,  to an initial value
\STATE Use Particle Smoother (PS) according to Algorithm \ref{alg:SMC} to compute $\hat{p}(y_{1:n}|\tilde{\theta})=L(\tilde{\theta})$ and sample $\widetilde{CU}(t_1:t_n)^{\tilde{\theta}}$ from $p(CU(t_1:t_n)|y_{1:n},\tilde{\theta})$
\FOR {$i=1$ to $M$}
	\STATE Sample $\tilde{\theta}^{*}$ from $Q(\tilde{\theta},.)$
	\STATE Use Particle Filter to compute $L(\tilde{\theta}^*)$ and sample $\widetilde{CU}(t_1:t_n)^{\tilde{\theta}^*}$ from $\hat{p}(CU(t_1:t_n)|y_{1:n},\tilde{\theta}^*) $
	\STATE Set $\tilde{\theta}=\tilde{\theta}^{*}$ and  $\widetilde{CU}^{\tilde{\theta}}(t_1:t_n)=\widetilde{CU}^{\tilde{\theta}^{*}}(t_1:t_n)$ with probability $1\wedge\frac{L(\tilde{\theta}^{*})Q(\tilde{\theta}^{*},\tilde{\theta})}{L(\tilde{\theta})Q(\tilde{\theta},\tilde{\theta}^{*})}$
	\STATE Record $\tilde{\theta}$ and $\widetilde{CU}^{\theta}(t_1:t_n)$
\ENDFOR
\end{algorithmic}
\end{algorithm}

Regarding the choice of $Q(.)$ in Algorithm \ref{alg:PMCMC}, we use a random walk Metropolis-Hastings algorithm in a transformed parameter space (log or logit) to ensure positivity. Each iteration of the MCMC algorithm requires an execution of the particle filter, which induces substantial computational cost if the importance sampling covariance matrix $\Sigma$ is ill-adapted. Adaptive approaches \citep{Roberts2009} can be used to tune $\Sigma$ but they require lengthy explorations of the target space. We propose to speed up this process by pre-exploration of a proxy posterior density $p^{EKF}(\theta|y)$ relying on a Gaussian approximation of the dynamic system and the Extended Kalman filter methodology \citep{Dureau2012}. A simple bootstrap version of the particle filter is used as it is not straightforward to consider data-driven transition proposals given complex observation regime of our model. Note that given the short length of the observed time series, simpler alternatives to PMCMC may perform reasonably well. For example, the resampling step can be omitted and the particle filter output can be used to approximate  $p(CU_{1:n}|y_{1:n},\theta)$. In our application however, it turns out that additional particles are needed for this approach, thus not offering a great reduction to the computational cost when compared to PMCMC. We therefore suggest the use of PMCMC as a robust computational tool that still does not requires a large amount of time in applications of this type. In order to meet with the computational requirements of the ensemble simulations at stake in the present article, and to facilitate future applications of the methodology we are proposing, the calculations were made using the \emph{SSM} inference package. This package, targeted to time series analysis via state space models, generates executables for likelihood optimisation and Bayesian inference algorithms in parallelisable C for any stochastic compartmental model expressed in a high-level modelling grammar (see \cite{Dureau2014} for more information).

\section{\label{EvalMeth}Evaluation methodology based on ensemble simulations}

Given the limited amount of information in available data (five prevalence observations, including initial conditions), it is very likely that the posterior output will be influenced substantially by the choice of CU priors and their parameters. In this section we explore the performance of the proposed inferential mechanism via simulation-based experiments designed to mimic the behaviour of datasets typically encountered in the context of application studied. Clearly, the approach of this paper heavily relies on the HIV infection model and the results will be dependent on its specification. We therefore set up the simulation experiments under the assumption that the model of Section \ref{subHIVmodel}, parameterised according to the priors of Section \ref{subseqPriors}, is correct. Focus is given on quantities related with the CU trajectories, under the different choices of Section \ref{subseqTrajPriors}, that can be estimated from the samples of the posterior distribution provided by the MCMC algorithms of Section \ref{subseqCompScheme}. We also provide some discussion regarding the static parameters appearing in the CU trajectory priors.

\subsection{Parameter of interest}
\label{sec:DCU}

By fitting each of the previously introduced models we obtain samples from the marginal posterior density $p^{meth}(CU(t)|y)$ ($meth\in\{dBR,dSigm,BM\}$). However, our interest mainly lies in the amplitude of the shift in CU between 2003 and April 2009 measuring the estimated increase in CU during the intervention, henceforth denoted by $\Delta CU$. The posterior draws of CU trajectories can be transformed to provide samples from the posterior of this parameter of interest. The samples can then be used to form an estimator $\hat{\Delta}CU^{meth}$ of $\Delta CU$ such as the posterior median of  $\hat{p}^{meth}(CU(t)|y)$. In what follows we explore the frequentist properties of this estimator derived from each of the trajectory priors.

It may also be of interest to assess the estimating capabilities, given the limited amount of data, for the hyperparameters of the various CU priors ($CU_0$, $\eta$, $r$, $m$, $t_{in}$ and $\sigma$). As it turns out there is information for some of them ($CU_0$, $\eta$, $t_{in}$), whereas some others are hard to estimate and are determined mostly by their prior ($r$, $m$ and $\sigma$). Nevertheless, from a subject matter point of view, interest lies mainly on $\Delta CU$, whereas the remaining quantities (in CU priors) can be regarded as nuisance parameters. Another appealing feature of $\Delta CU$ is that it appears in all models and therefore provides an omnibus quantity for comparison. Hence, inference properties of these parameters ($CU_0$, $\eta$, $r$, $m$, $t_{in}$ and $\sigma$) are only studied indirectly through inference properties of $\Delta CU$.

\subsection{Measures of performance}

The performance of each estimator $\hat{\Delta} CU^{meth}$ in estimating $\Delta CU$ is evaluated from the following criteria (where $L=1000$, the number of simulations):

\begin{equation}
\begin{array}{rl}
Bias^{meth} &= \frac{1}{L}\sum_i (\hat{\Delta}CU(t)^{meth} - \Delta CU(t) )\nonumber\\
MSE^{meth} &= \frac{1}{L}\sum_i (\hat{\Delta}CU(t)^{meth} - \Delta CU(t) )^2\nonumber\\
Std^{meth} &= \sqrt{MSE^{meth}  - (Bias^{meth} )^2 }\nonumber\\
\end{array}
\end{equation}

In addition to the quantities above we are also interested in assessing the discriminative ability of each model in detecting increases in CU. Focus is given to increases in CU that are at least as high as a pre-specified threshold T, which reflects the minimum practical increase. When analysing the data a researcher may decide that CU did increase more than T if the value of the estimator $\hat{\Delta} CU^{meth}$   is higher than a user-specified threshold $t$. Each decision mechanism may lead to different types of error and is therefore associated with a particular sensitivity and specificity. More specifically we can define the true and false positives in the following way

\begin{itemize}
\item Sensitivity (true positives rate) for $t$: $\frac{\sharp(\hat{\Delta} CU^{meth}>t,\Delta CU>T)}{\sharp(\Delta CU>T)}$
\item Specificity (1 - false positives rate) for $t$: $\frac{\sharp(\hat{\Delta} CU^{meth}<t,\Delta CU<T)}{\sharp(\Delta CU<T)}$
\end{itemize}

We proceed by first reporting sensitivities and specificities corresponding to the case of  $t=T$. This corresponds to saying that  $\Delta CU$  is higher than $T$ if its estimator is higher than $T$. We then use a range of different $t$'s and obtain the sensitivity-specificity pair that corresponds to each  of them. A lower detection threshold $t$ will increase the sensitivity of the method, but it also increases the risk for false positives, and vice versa. These pairs are  combined to form the Receiver Operating Characteristics ROC curve by plotting sensitivity versus 1-specificity. The area under the ROC curve (AUC)  provides an overall measure of discriminatory power as it reflects the probability of correctly classifying a randomly chosen positive instance as higher than a randomly chosen negative one (Fawcett, 2006). For example, an AUC value of 50\% indicates no power (i.e random choice)  This detailed procedure is repeated to assess the ability to detect two different levels of increase in CU, with $T$ set to $20\%$, $30\%$ and $40\%$ respectively.

\subsection{Simulation procedure}

The performance of the estimators derived from the different models is measured using a set of simulated experiments where CU trajectories are sampled from a given growth curve model ($dSigm$), and parameters from $\theta_{i.c.}$ and $\theta_{tr.}$  are sampled following their prior distributions. To maximise the utility of this test procedure for future application of this methods to help evaluate Avahan in different districts (manuscript in preparation), only plausible and realistic CU trajectories are considered: cases with prevalence in 2010 between 2\% and 40\% and with CU shifts that occurred after 1995. Furthermore, the test trajectories have been sampled so that  $\Delta CU$ regularly spans the $[0;0.9]$ interval. Alternatively, a second set of simulations has been generated based on step-wise CU trajectories (constant CU up to $t_{in}$, and then constant CU after $t_{in}$) to explore the impact of model misspecification and to assess the sensitivity of results to the underlying model used for simulations ($dSigm$).

For each of these experiments, an epidemic is simulated to provide observations  $(y_i^{sim})$ replicating the observation framework applied in Mysore: three prevalence estimates among female sex workers and one among clients, concentrated during the period of the intervention. From these observations, the MCMC algorithm is applied to each method $meth$ to sample from $p(CU^{meth}(t_1:t_n)|y^{sim}_{1:n})$. Then, given the posterior CU samples the estimators $\hat{\Delta} CU^{meth}$ can be computed, and compared to their true counterparts $\Delta CU$ by calculating the measures of performance of the previous subsection.

%

\section{\label{Results}Results}

\subsection{\label{subseqResults}Comparison of the CU trajectory models from ensemble simulations}

This section contains the results of the simulation experiments that were conducted as described in the section 3.3. In all cases each model is fit to each of the $1000$ simulated datasets and the estimator is formed via the posterior median at the points of interest. Various frequentist properties of these estimators are then assessed such as bias, the standard deviation and the MSE. Moreover, in Table \ref{DistProps}, we examine the ability of each model to classify the simulated instances of the $\Delta CU$ parameter, and assess the risk of overstating versus understating. In other words, we assess the ability of model-driven estimators to address questions such as 'was the shift in CU during the intervention over 0.2, 0.3, or 0.4?', by looking at their sensitivity and specificity as well as the resulting AUC.

\begin{figure}
\centering
\makebox{\includegraphics[scale=0.7]{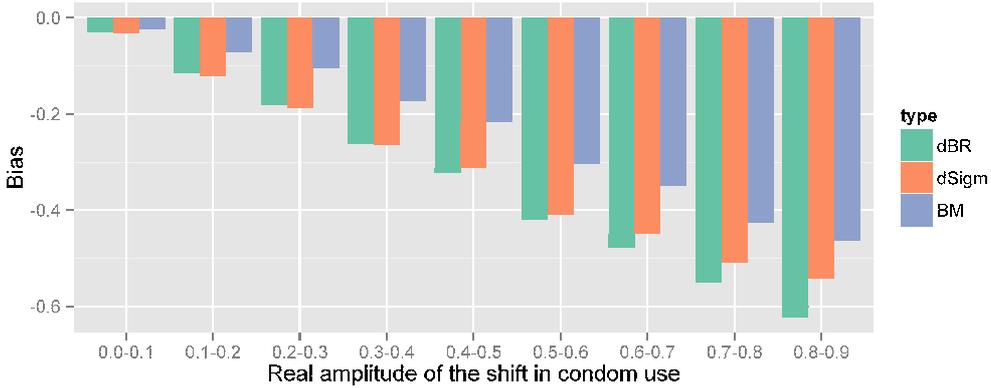}}
\caption{Bias of each model as a function of the true amplitude of the shift in condom use, estimated from 1000 simulations.}
\label{fig:Bias}
\end{figure}

We begin with results corresponding to cases where the data were simulated from the dSigm model. Table \ref{FreqPropsdSigm} reports bias, standard deviation and MSE for each model-driven estimator. It suggests that all models underestimate $\Delta CU$. More precisely, the dBR model tends to strongly understate the shift in amplitude, by 0.33 in average. The biases of the dSigm model is only slightly smaller (-0.31), and optimal results are obtained with the Brownian Motion model (-0.26). Similarly, in terms of MSE, the performance of the BM model is better. Figure \ref{fig:Bias} plots the bias against the true value of $\Delta CU$ for each model and reveals an increasing association. Another interesting note is that the ranking of the different models is consistent across the entire range of $\Delta CU$ values. If, for example, the true shift is between 40\% and 50\%, it is on average underestimated by 0.21 with the best method (BM) and more than 0.30 points with the dBR and dSigm methods.

\begin{table}
\caption{\label{FreqPropsdSigm}Frequentist properties of the different estimators of the amplitude of the shift in condom use during the intervention, estimated from 1000 simulations from the dSigm model}
\centering
\fbox{%
\begin{tabular}{|cccc|}
\hline
 & \emph{Deterministic}   & \emph{Deterministic}  & \emph{Brownian} \tabularnewline
 & \emph{Bertalanffy-} & \emph{empirical}   & \emph{motion}\tabularnewline
 & \emph{Richards}& \emph{sigmoid}  & \tabularnewline

Bias & -0.33 & -0.31 & -0.26\tabularnewline
Error standard deviation & 0.21  & 0.20  & 0.19\tabularnewline
Mean Squarred Error (MSE) & 0.15 & 0.14 & 0.10\tabularnewline
\hline
\end{tabular}}
\end{table}

The bias in estimating $\Delta CU$ under each of the CU priors can be attributed to a large extent to the prior implied by each formulation on $\Delta CU$. As mentioned in Section \ref{subseqPriors} the BM approach results in a symmetric prior on $\Delta CU$ that is centered around 0 with 2.5\% and 97.5\% points at $\pm 0.6$ respectively. The posterior median is therefore pulled towards zero, thus resulting in conservative estimates. The amount of shrinkage depends on the upper limit of the Uniform prior on $\sigma$. The corresponding priors under the dBR and dSigm formulations result in priors for $\Delta CU$ that put more mass around zero and this heavily depends on the priors placed on $k$ and $m$. It would perhaps be interesting in the future to consider alternative priors for $k$ and $m$ to reduce the bias in $\Delta CU$. Regarding the dBM model, the Uniform prior assigned on $\sigma$ appears to work reasonably well.

\begin{figure}
\centering
\makebox{\includegraphics[scale=0.8]{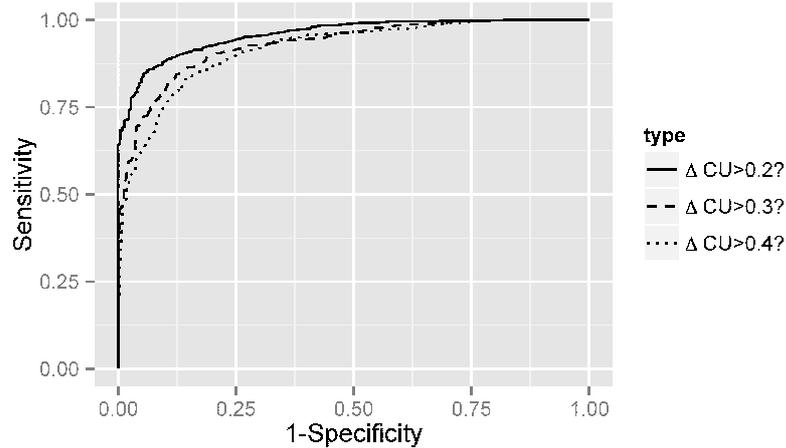}}
\caption{ROC curve when testing for $\Delta CU>0.2$, $\Delta CU>0.3$  and $\Delta CU>0.4$, under Brownian motion trajectory  prior. These curves were estimated from 1000 simulations. Very similar shapes are obtained for the alternative trajectory priors.}
\label{fig:ROC}
\end{figure}

Table \ref{DistProps} suggests that all estimators based on the median of the posterior density of $p(\Delta CU |y_{1:n})$ have good distinguishing power: the AUC is between 0.87 and 0.93 in all cases. In line with the results of Table \ref{FreqPropsdSigm}, the estimates provided by the BM model achieve better sensitivity (53\%, 29\% and 14\%) than the other models (between 0\% and 29\%). The strong negative bias observed in Table \ref{FreqPropsdSigm} is also reflected by an optimal 100\% specificity of the estimators. The performance, in terms of sensitivity, decreases as the level of $\Delta CU$ increases. Figure \ref{fig:ROC} also depicts the ROC curve for the dBM model, obtained as described in section 3.3.

\begin{table}
\caption{\label{DistProps}General distinctive power (AUC) of the median estimator of the shift, and specific sensitivity and specificity when answering: is the shift in CU during the intervention stronger than 0.2?  than 0.3? than 0.4? These quantities were estimated over 1000 simulations from the dSigm model.}
\centering
\fbox{%
\begin{tabular}{|ccccc|}
\cline{3-5}
\multicolumn{1}{c}{} & & \emph{Deterministic}   & \emph{Deterministic}  & \emph{Brownian} \tabularnewline
\multicolumn{1}{c}{} &  & \emph{Bertalanffy-}  & \emph{Empirical} &  \emph{motion}\tabularnewline
\multicolumn{1}{c}{} &  & \emph{Richards} & \emph{Sigmoid} &  \tabularnewline
\multirow{3}{*}{$\Delta CU>0.2?$} & AUC & 0.93 & 0.93 & 0.92\tabularnewline
 & Sensitivity & 26\%  & 29\% & 53\%\tabularnewline
 & Specificity & 100\%  & 100\% & 100\%\tabularnewline
 \multirow{3}{*}{$\Delta CU>0.3?$} & AUC & 0.92 & 0.92 & 0.89\tabularnewline
 & Sensitivity & 5\%  & 14\% & 29\%\tabularnewline
 & Specificity & 100\%  & 100\% & 100\%\tabularnewline
\multirow{3}{*}{$\Delta CU>0.4?$} & AUC & 0.90 & 0.89 & 0.87\tabularnewline
 & Sensitivity & 0\% &  7\%  & 14\%\tabularnewline
 & Specificity & 100\%  & 100\%  & 100\%\tabularnewline
\hline
\end{tabular}}
\end{table}

In an additional simulation experiment we examine the frequentist properties of the models in the case where underlying CU trajectories were generated from a step function. As shown in Table \ref{FreqPropsstep} a very similar picture is obtained with the BM model possessing the smallest bias and MSE, while dSigm being slightly better than dBR.

\begin{table}
\caption{\label{FreqPropsstep}Frequentist properties of the different estimators of the amplitude of the shift in condom use during the intervention, estimated from 1000 simulations from step-wise CU trajectories (constant CU until $t_{in}$, and constant CU after $t_{in}$.}
\centering
\fbox{%
\begin{tabular}{|cccc|}
\hline
 & \emph{Deterministic}   & \emph{Deterministic}  & \emph{Brownian} \tabularnewline
 & \emph{Bertalanffy-} & \emph{empirical}   & \emph{motion}\tabularnewline
 & \emph{Richards}& \emph{sigmoid}  & \tabularnewline

Bias & -0.35 & -0.33 & -0.26\tabularnewline
Error standard deviation & 0.21  & 0.20  & 0.18\tabularnewline
Mean Squarred Error (MSE) & 0.16 & 0.15 & 0.10\tabularnewline
\hline
\end{tabular}}
\end{table}

Overall, the results presented in these tables provide an informative qualitative assessment for the ability of the different models to capture $\Delta CU$ from limited prevalence data on an important and diverse set of likely scenarios (1000 experiments with true amplitude in CU shift spanning between 0 and 0.9). First of all MSE and AUC figures suggest that although the number of prevalence observations is low and some elements of the transmission process are uncertain, it is still possible to extract information on our time varying parameter and provide estimates of the amplitude of the shift in CU during the intervention. Furthermore, there seems to be a possibility to control the risk of overstating these quantities by analysing the outputs of the three models that offer different levels of compromise between sensitivity and specificity. Thus, although the procedure may fail to identify some shifts in CU, there seems to be some reassurance that once a shift is detected it will most likely be true. In other words the obtained model-driven estimators seem to be conservative, yet robust on detecting CU trends.

The two models with the higher overall performance, BM and dSigm, are quite different in nature: the dSigm trajectories are smooth, whereas under the Brownian motion prior they are non-differentiable. Hence, the choice between the two models can also be based on prior beliefs of the researcher regarding the smoothness of the CU trajectories.

\subsection{Application: what can we infer on the trajectory of CU in Mysore?}

\begin{figure}
\centering
\makebox{\includegraphics[scale=0.6]{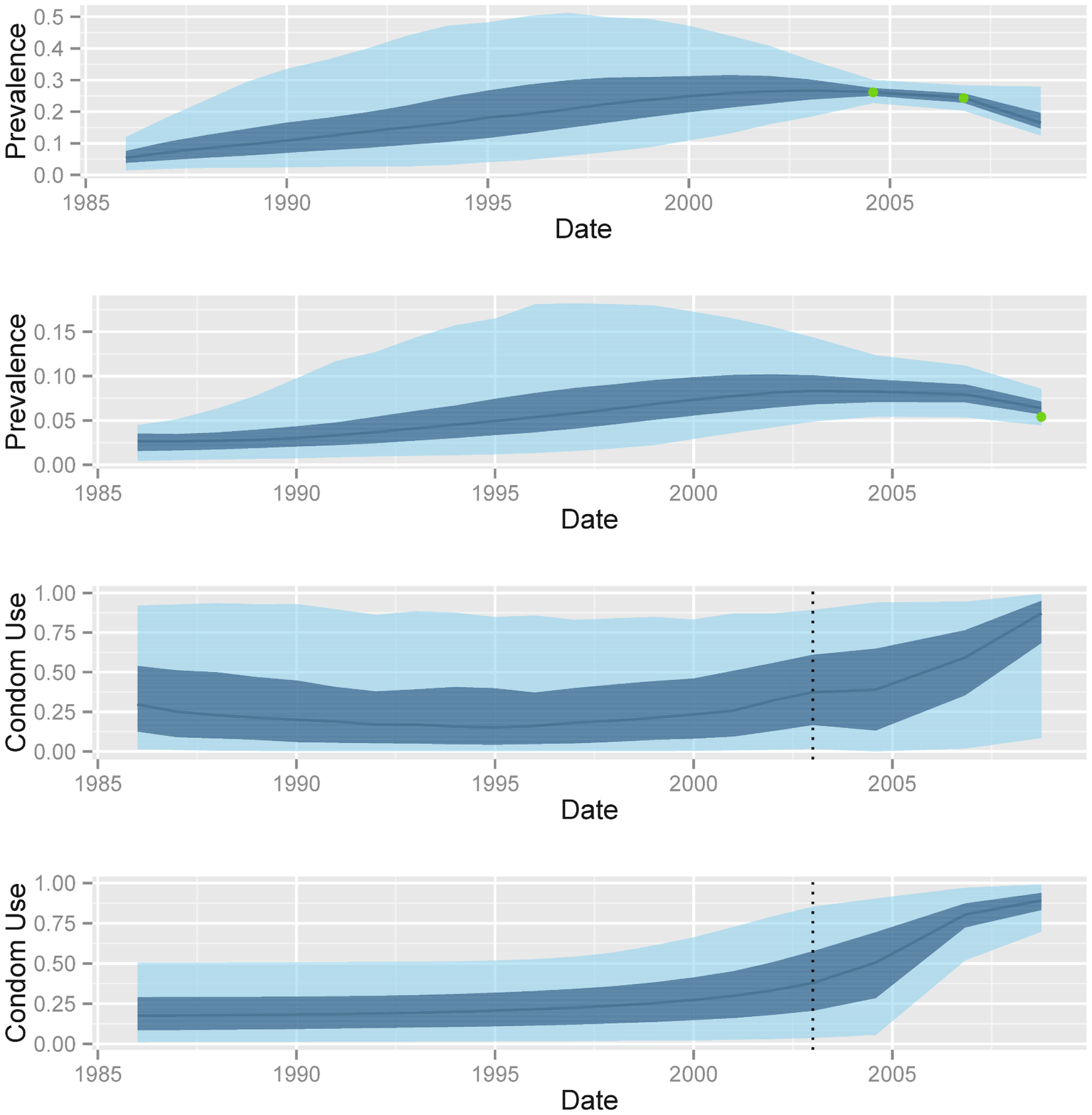}}
\caption{Estimates obtained for Mysore district.  \protect \\
a) reconstructed prevalence trajectory among female sex workers when condom use modelled with Brownian motion  \protect \\
b) reconstructed prevalence trajectory among clients when condom use modelled with Brownian motion  \protect \\
c) reconstructed condom use trajectory when modelled with Brownian motion  \protect \\
d) reconstructed condom use trajectory when modelled with deterministic Sigmoid}
\label{fig:Trajs}
\end{figure}

Mysore is one of the districts targeted by the Avahan intervention, and Avahan was the first HIV prevention intervention in this region. Four HIV prevalence estimates have been obtained between 2003 and 2009, three among female sex workers, and one among clients. Results from the inference procedure using a Brownian motion model are shown in Figure \ref{fig:Trajs}, suggesting a strong impact of the intervention. The purpose of this paper was to determine what level of increase of CU between 2003 and 2009 can be inferred while controlling the risk of overstating it. As it was shown in section \ref{subseqResults}, dSigm models could provide a good alternative to the BM formulation. Hence, we also present here results obtained with this model for the Mysore dataset (see Figure \ref{fig:Trajs}). An interesting point to note here is the lower posterior uncertainty around the CU trajectory in the case of the dSigm model. This is in line with our expectations given the fact that this prior imposes a particular growth structure and therefore introduces additional information. Loosely speaking this may be viewed as a bias-variance tradeoff. The choice of each CU formulation is likely to affect the inference procedure; therefore it is useful to explore its implications, for example via the simulation experiments of the previous section. Table \ref{MysoreTable} shows the estimates of $\Delta CU$ for each of the three presented models. The results indicate a positive increase in all cases. In particular, for the BM and dSigm models the corresponding posterior medians are 0.54 and 0.49 while the 95\% credible intervals are [0.06;0.95] and [0.09;0.97] respectively.

A stronger conclusion regarding a lower bound for the CU shift between 2003 and 2009 can be made by comparing the posteriors medians to the results of Table  \ref{DistProps}. If the underlying set of simulations is to be considered realistic, an argument in favour of a CU increase being at least 0.4 can be made. Since the posterior medians are more than 0.4 under both BM and dSigm models (.54 and .49 respectively), Table  \ref{DistProps} suggests that a statement for $\Delta CU>0.4$ will be correct with probability given by the specificity of each model (100\% in both cases). While being more informative than the credible intervals obtained directly from the posterior densities (over 0.06), these numbers are heavily dependent on the assumption that the simulations of Section \ref{EvalMeth} provided an adequate approximation of the reality. Finally, Figure \ref{fig:Trajs} and Table \ref{MysoreTable} show that the results obtained from the deterministic Sigmoid and Brownian motion models strongly coincide: they suggest that CU was stable over the 1985-2003 period, remaining below 0.5, sharply increased between 2003 and 2007, and kept growing until 2009.

\begin{table}
	\caption{\label{MysoreTable}Estimates of the change in CU in Mysore between 2003 and 2009.}
\centering
\fbox{%
\begin{tabular}{|ccccc|}
\multicolumn{1}{c}{} &  & \emph{Posterior} & \emph{Posterior} & \emph{$95\%$ credible} \tabularnewline
\multicolumn{1}{c}{} &  & \emph{mean} & \emph{median} & \emph{interval}\tabularnewline
\multirow{3}{*}{$\Delta CU$} & Deterministic Bertalanffy-Richards & 0.30 & 0.29 & $[0.09;0.54]$\tabularnewline
 & Deterministic Sigmoid & 0.48 & 0.49 & $[0.09;0.87]$\tabularnewline
 & Brownian motion & 0.52 & 0.54 & $[0.06;0.95]$\tabularnewline
\hline
\end{tabular}}
\end{table}

Overall, the results of this section support and strengthen the findings of various papers\citep{Boily2007,Deering2008,Boily2008,Lowndes2010,Pickles2010,Boily2013,Pickles2013} which relied on very informative prior information from self reported estimates of condom use to assess the impact of the Avahan initiative on HIV infection averted. The implications of our analysis via the proposed methodology of the paper is the ability to draw inference on the space of CU trajectories without informing the model about the timing of the Avahan initiative or relying on other sources of data such as FSWs surveys of condom distribution data.

\section{\label{Discussion}Discussion}

In this article, we presented a Bayesian approach to draw conclusions regarding the evolution of time-varying behavioural parameters in the context of HIV such as CU among FSWs. Inference can be based on prevalence estimates while a substantial amount of information from additional sources can be incorporated via prior distributions. In order to describe the behaviour of CU trajectories we introduced three different formulations based on Brownian motion and growth curves such as the generalised Bertalanffy Richards and empirical sigmoid models. To our knowledge, these formulations are new in this context. The presented computational framework allows estimation of CU trajectories as well as functionals thereof, using advanced MCMC methods and following ideas of \cite{Dureau2012, Cazelles1997, Rasmussen2011}. Nevertheless, in comparison to these approaches, the problem of evaluating the Avahan intervention by estimating its impact on CU from prevalence estimates is of additional difficulty due to the limited amount of information; the application to Mysore district was based on three observations of prevalence among FSWs and one among clients, plus hypothesis on the initial value of prevalence in 1985. Various simulation experiments were conducted in order to assess the validity of the procedure, examining the frequentist properties of the underlying estimators and the ability of the model to avoid overestimation via conducting ROC analysis. The evidence from the simulation experiments is encouraging, suggesting that the approach can be used in this context for making conservative estimates of changes in CU both with the Brownian motion and the deterministic sigmoid trajectory priors. However, the overall performance is bound to depend on the deterministic HIV infection model which was parameterised based on a substantial amount of prior information, as in \cite{Pickles2010}, as well as on assumptions such as the very low HIV prevalence in 1985. Most of the prior information utilised in this study was obtained from additional data sources (IBBAs). In the presence of all these data sources, it would be interesting to consider and contrast a joint inferential scheme through an evidence synthesis framework in the spirit of \cite{Goubar2008,Presanis2011}.

While the representation of HIV transmissions in this paper is simpler in behavioural terms in comparison with the model presented in \cite{Pickles2010}, the model is enriched as it explores the CU trajectories space rather than working with three pre-determined scenarios. Nevertheless, there are reasons for a potential overestimation of the shift amplitude in this simpler model as coinfection with other sexually transmitted diseases were ignored (although higher transmission probability per unprotected act were allowed to compensate for the latter), and no acute phase was considered. However, diffusion driven models aim at capturing and compensating for structural mis-specifications while capturing the main dynamics of the system and have been shown here to provide conservative estimates. Overall it may be viewed as a different and complementary choice in the trade-off between richness and tractability of the model compared to \cite{Pickles2010}. Lastly, this approach relies on the hypothesis that changes in transmission probabilities are solely related to changes in CU, ignoring for example potential changes in the frequency of commercial sex partnerships. This choice can be motivated by the strong focus of the Avahan intervention on prevention measures and the relative stability in the frequency of commercial sex exhibited by the series of cross-sectional bio-behavioural surveys that were conducted during the period of the intervention.

\section*{Acknowledgements}
This research was funded by the Bill \& Melinda Gates Foundation. The views expressed herein are those of the authors and do not necessarily reflect the official policy or position of the Bill \& Melinda Gates Foundation, the London School of Economics, the London School of Hygiene \& Tropical Medicine, or Imperial College London. Many thanks to Bernard Cazelles and to the Eco-Evolutionary Mathematics team of UMR 7625, Ecology \& Evolution for use of High Performance Computing facilities. The authors will also like to thank the Associate Editor and the Referees for their thorough and constructive review.

\bibliographystyle{chicago}
\bibliography{Biblio}

\end{document}